\documentclass[conference]{IEEEtran}
\usepackage{graphicx}
\usepackage{subfig}
\usepackage{amsmath}
\usepackage{hyperref}
\usepackage[color=green!10]{todonotes}
\newcommand{\exc}{\textnormal{E}}     
\newcommand{\external}{\textnormal{X}}     

\newcommand{\Epop}{\mathcal{E}} 
\newcommand{\Ipop}{\mathcal{I}} 

\newcommand{\effweight}{W}     
\newcommand{\effdelay}{d}   

\newcommand{\wEE}{\effweight_{\exc\exc}}

\newcommand{\wEX}{\effweight_{\exc\external}}
\newcommand{\dEX}{\effdelay_{\exc\external}}
\newcommand{\Xpop}{\mathcal{X}} 

\newcommand{\dap}{\textnormal{dAP}}
\newcommand{\Seq}{\mathcal{S}}

\def\BibTeX{{\rm B\kern-.05em{\sc i\kern-.025em b}\kern-.08em
    T\kern-.1667em\lower.7ex\hbox{E}\kern-.125emX}}

\title%
{Unsupervised Learning of Spatio-Temporal Patterns in Spiking Neuronal Networks}

\author{\IEEEauthorblockN{Florian Feiler}
\IEEEauthorblockA{\textit{Peter Gr\"unberg Institute (PGI-15)} \\
 \textit{J\"ulich Research Centre}\\
Germany \\
f.feiler@fz-juelich.de}
\and
\IEEEauthorblockN{Emre Neftci}
\IEEEauthorblockA{\textit{Peter Gr\"unberg Institute (PGI-15)} \\
 \textit{J\"ulich Research Centre}\\
\textit{RWTH Aachen}\\
Germany \\
e.neftci@fz-juelich.de}
\and
\IEEEauthorblockN{Younes Bouhadjar}
\IEEEauthorblockA{\textit{Peter Gr\"unberg Institute (PGI-15)} \\
\textit{Peter Gr\"unberg Institute (PGI-7)}\\
 \textit{J\"ulich Research Centre} \\
Germany \\
y.bouhadjar@fz-juelich.de}
}

\begin{document}

\maketitle
\begin{abstract}
The ability to predict future events or patterns based on previous experience is crucial for many applications such as traffic control, weather forecasting, or supply chain management.
While modern supervised Machine Learning approaches excel at such sequential tasks, they are computationally expensive and require large training data.
A previous work presented a biologically plausible sequence learning model, developed through a bottom-up approach, consisting of a spiking neural network and unsupervised local learning rules.
The model in its original formulation identifies only a specific type of sequence elements composed of synchronous spikes by activating a subset of neurons with identical stimulus preference.
In this work, we extend the model to detect and learn sequences of various spatio-temporal patterns (STPs) by incorporating plastic connections in the input synapses.
We showcase that the model is able to learn and predict high-order sequences. We further study the robustness of the model against different input settings and parameters.
\end{abstract}


\section{Introduction}
Learning and processing sequences of events is central to human cognition. While supervised machine learning methods achieve astounding results at sequential processing, they provide few insights into underlying biological mechanisms, are energy-intensive, and are prone to failure in noisy or incomplete environments. Borrowing solutions from biological neuronal networks is a promising pathway to overcome such limitations.
Recent studies demonstrated the potential of dendritic computations for a sequence prediction task \cite{bouhadjarSequenceLearningPrediction2022, asabukiNeuralCircuitMechanisms2022}.
The work in \cite{bouhadjarSequenceLearningPrediction2022} developed a biologically inspired sequence learning and prediction model. It learns to predict complex sequences in an unsupervised, continuous manner using biological, local learning rules.  
The model identifies the sequence elements through a prewired mechanism, i.e., the presentation of a sequence element activates a specific subpopulation of neurons.
In this work, we aim to extend the model to learn in an unsupervised manner naturalistic sequences composed of a stream of spatio-temporal patterns (STPs).
We study the model's robustness to noise and hyperparameter choice, thereby exploring the potentials and limitations of dendritic processing for sequence learning and processing.

The network achieves online, bio-plausible, and unsupervised sequence stimulus detection using spike time-dependent plasticity (STDP) and competition via inhibition. 

\section{Model}
\begin{figure}[htbp]
    \centering
    \subfloat{\includegraphics[width=1\linewidth]{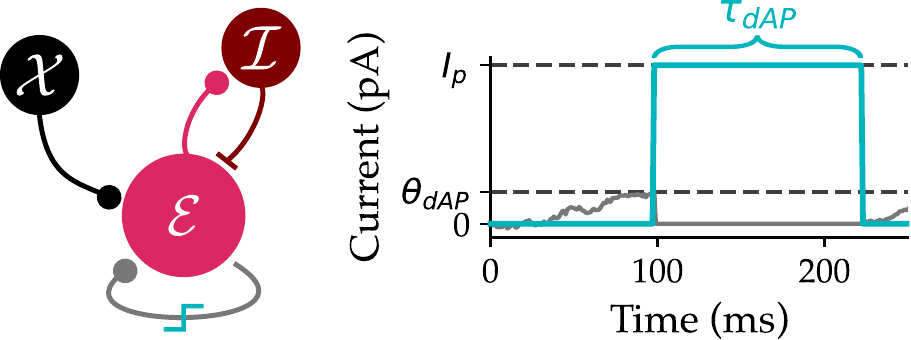}}
    \caption{Network overview and illustration of the dendritic action potential (dAP). (Left) The excitatory population $\Epop$ receives input from external spike sources $\Xpop$, inhibition via the inhibitory population $\Ipop$, and is recurrently connected to itself. All populations are sparsely connected. (Right) Recurrent input is integrated at the dendrite. Neurons are predictive if currents cross the dAP threshold $\theta_\dap$. Predictive neurons are depolarized by a strong positive input current $I_\textnormal{p}$ for a constant time of $\tau_\dap$ (light blue).}
    \label{fig:network_sketch}
\end{figure}

The network consists of three sparsely recurrently connected neuron populations: excitatory neurons ($\Epop$), inhibitory neurons ($\Ipop$), and external spike sources ($\Xpop$), thereby constituting a recurrent EI network (Fig.~\ref{fig:network_sketch}). Synaptic delays between populations are constant, except for $\dEX$. A non-linear activation function mediates dendritic prediction (Fig.~\ref{fig:network_sketch}).

\begin{figure}[htbp]
\centerline{\includegraphics[width=1.0\linewidth]{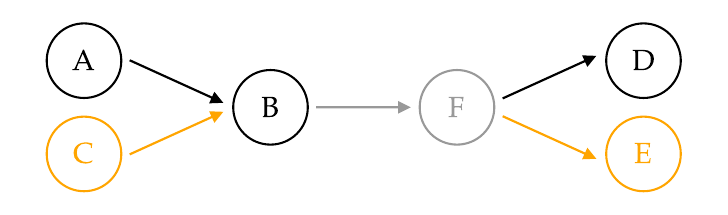}}
    \caption{High-order sequences of different complexities presented to the network. The more items both sequences overlap, the longer the network has to memorize the initial item to predict the final one. We consider two sequence sets throughout this study:  $\Seq_3 = \left\{\textit{ABD}, \textit{CBE}\right\}$ and $\Seq_4 = \left\{\textit{ABFD}, \textit{CBFE}\right\}.$}
    \label{fig:input_sequences}
\end{figure}

Inputs are composed of sequences $\Seq$ of STPs, abstractly described by Latin letters $A, B,\ldots, Z$. For each run, two partially overlapping and therefore high-order sequences are considered and presented to the network in an alternating fashion (Fig.~\ref{fig:input_sequences}). To increase the bio-plausibility of the setup and test the model's robustness to noise, we introduce Poisson-generated noise  between sequence stimuli and, for a prolonged duration, between the two sequences.
Noise duration and firing rates are defined in relation to sequence item characteristics (see Results section for further details).

\begin{figure}[!htbp]
    \centerline{\includegraphics[width=1.0\linewidth]{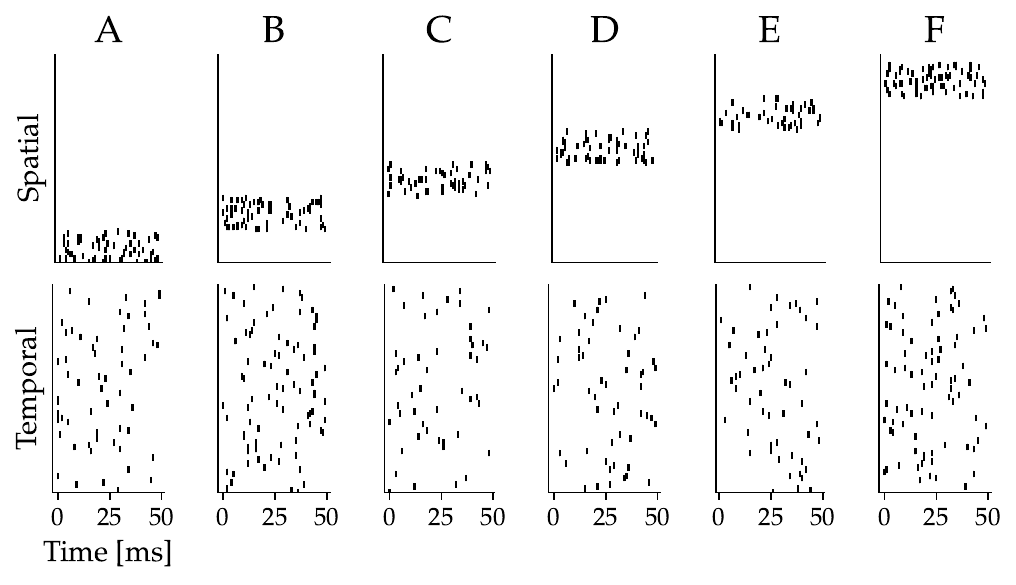}}
    \caption{Exemplary sequence items for spatial and temporal encoding for the sequence set $\Seq_4$. All items are generated from a Poisson process and share a duration of $t_\textnormal{item} = 50\,\textnormal{ms}$. During learning, patterns are shown repeatedly, with distraction noise of the same duration between sequence items. A fraction of 50\% of all spikes is shown for visualization purposes. }
    \label{fig:input_scatter}
\end{figure}

The overlapping sequence elements introduce ambiguity. To resolve these ambiguities, the network must form context-dependent representations for the overlapping elements. The context is described by the sequence currently shown to the network. A presentation of a sequence element consists of a mapping to a specific STP.
STPs can be divided into two central characteristics: purely spatial and purely temporal pattern encoding. Spatial patterns are characterized by the occupied subset of input channels in $\Xpop$. Temporal patterns occupy all channels but differ in their exact spike time throughout the channels (Fig.~\ref{fig:input_scatter}). 

Excitatory neurons are modeled as leaky-integrated firing (LIF) neurons, which are characterized by having three input channels: external stimulation, inhibition via $\Ipop$, and recurrent excitatory synapses. 
Synapses from external sources have weights $\wEX$ and delays $\dEX$, which are both initialized from normal distributions, with the plastic weights following STDP type of plasticity. Inhibitory synapses receiving input from LIF neurons have a constant weight and delay.

Inhibition enforces competition between excitatory neurons, leading to item-specific selectivity.
In addition to the homeostatic plasticity, this results in non-overlapping excitatory responses. Recurrent weights $\wEE$ are initialized from a random distribution and are updated according to a time-restricted STDP learning rule \eqref{eq:dap_stdp}. The restriction ensures taking spiking activity during the presentation of the previous item into account while excluding responses to the current and second-to-last item, thereby allowing learning sequence item transitions and preventing direct context transfer from earlier sequence elements. 

Excitatory neurons can be in one of two states: predictive or not predictive. Neurons are considered predictive if currents from recurrent input are strong enough to cross the dendritic action potentials (dAPs) threshold $\theta_\dap$. Then, for a constant time of $\tau_\dap$, neurons are strongly depolarized by receiving a positive input current $I_\textnormal{p}$. After that phase, neurons return to the default non-predictive state.

The dAPs play an important role in learning context-dependent representations of overlapping items, especially since they can sustain activity while stimulus is absent. Predictive neurons are expected to fire earlier and therefore suppress the activity of the non-predictive neurons. Thereby, overlapping items develop two context-dependent sets of predictive neurons, which in turn allow context-dependent activation of the next element.

Simulation is split into two phases. During the first phase, we only allow plastic $\wEX$ connections to grow. This phase supports assembly formation and neuron specialization. During the second phase, synaptic weights $\wEX$ are frozen and only recurrent connections $\wEE$ are plastic.
With the onset of the second phase, inhibition levels are increased to support the creation of context-dependent predictive neurons. To further strengthen the context-dependent activation, weight updates are controlled by homeostatic control, which constrains neurons to be predictive only in one context:

\begin{equation}
    \begin{aligned}
        \Delta w_{\exc\exc}^+ &= 
        \begin{cases} 
            \lambda^+ \left( z_\textnormal{t} - z_\textnormal{p} \right) w_{\exc\exc} \,\textnormal{tr}_\textnormal{pre} & \text{if } \textnormal{dt}_\textnormal{min} \leq \Delta t \leq \textnormal{dt}_\textnormal{max}\\
            0 & \text{else }
        \end{cases}
        \\
        \Delta w_{\exc\exc}^- &= \lambda^-,
    \end{aligned} \label{eq:dap_stdp}
\end{equation}

where $z_\textnormal{t}$ is the target firing rate for homeostatic control of the current firing rate $z_\textnormal{p}$, $\textnormal{tr}_\textnormal{pre}$ the presynaptic trace, and positive updates occur only if the time gap $\Delta t$ between pre- and postsynaptic spike is between sensible bounds.

For visualization and evaluation, excitatory neurons are grouped by their maximum average response per item. Those selective groups are called assemblies and are denoted by their corresponding item stimulus, e.g., $\Epop_\textnormal{B}$. We measure the average number of predictive neurons in each assembly. The item corresponding to the assembly with the most predictive neurons is predicted. Prediction accuracy is evaluated only for the last item in each sequence.  If not explicitly mentioned, all experiments are carried out $20$ times.

In the current implementation, network size and simulation time scale proportionally with the number of items in the sequence sets. 
Implementation details and used hyperparameters can be found at \href{https://github.com/ffeiler/unsupervised_sequence_learning_stp_snn}{github}.

\section{Results}

\subsection{Spatial Encoding}
\begin{figure}[htbp]
    \centering
	\subfloat{\includegraphics[width=0.49\linewidth]{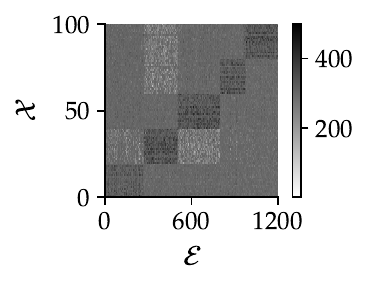}}
	\hspace{0.01cm}
	\subfloat{\includegraphics[width=0.49\linewidth]{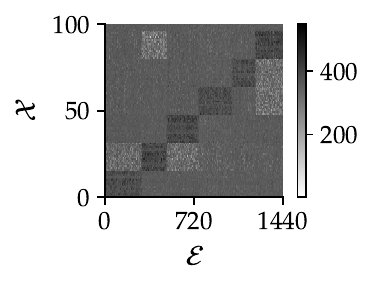}}
    \caption{Exemplary input weights $\wEX$ after simulation for sequence set $\Seq_3$ (left) and $\Seq_4$ (right). Unrealized synapses are plotted with the default value (gray) to not confuse them with depressed (white) or potentiated weights (black).}
    \label{fig:connectivity_EX}
\end{figure}

During the first simulation phase, after the network is repetitively exposed to the external stimuli, it learns non-overlapping sequence item-specific assemblies (Fig.~\ref{fig:connectivity_EX}).

\begin{figure}[htbp]
    \centering
	\subfloat{\includegraphics[width=1\linewidth]{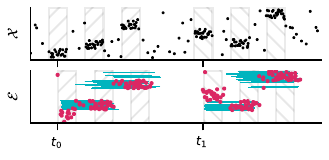}}
    \caption{Exemplary network activity after learning for thee spatially encoded sequence set $\Seq_3$.
    Input spikes (black) arrive with a synaptic delay at excitatory neurons and can trigger excitatory spikes (red). Neurons are in predictive mode and depolarized if a dendritic action potential is generated (light blue). $t_0$ and $t_1$ indicate start times of both sequences in $\Seq_3$, corresponding to one context each (hatched items). For visualization purposes, only 25$\%$ of all spikes are shown. Excitatory neurons are sorted by item selectivity, and inhibitory activity is not shown.}
    \label{fig:example_network_activity}
\end{figure}

\begin{figure}[htbp]
    \centering
	\subfloat{\includegraphics[width=0.49\linewidth]{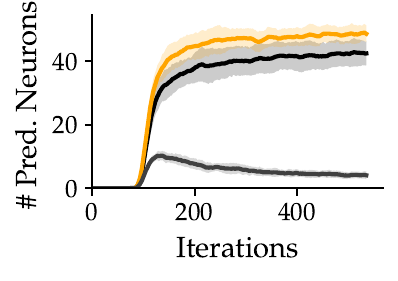}}
	\hspace{0.01cm}
	\subfloat{\includegraphics[width=0.49\linewidth]{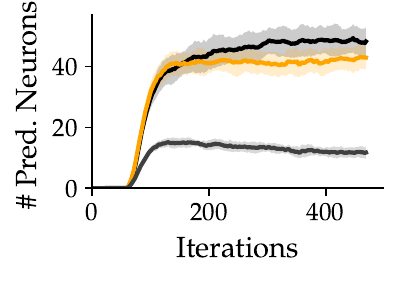}}
    \caption{Number of predictive neurons for context 1 (black) and context 2 (orange), and the overlap between both sets (gray) for the last sequence item in sequence sets $\Seq_3$ (left) and $\Seq_4$ (right). Curves and error bands indicate a moving average ($k=10$) of mean and one standard deviation.}
    \label{fig:nb_daps_all_seqlens}
\end{figure}

In the second phase, the network learns the sequence prediction task. The network activity after training exhibits context-dependent dAP predictions, which in turn influence excitatory responses (Fig.~\ref{fig:example_network_activity}). Initially, we observe a rapid forming of predictive neurons, but with a certain overlap in the activity patterns of two contextual inputs (Fig.~\ref{fig:nb_daps_all_seqlens}). Over time, homeostasis and increased inhibition levels lead to a decrease in overlaps. 

\begin{figure}[htbp]
    \centering
	\subfloat{\includegraphics[width=0.49\linewidth]{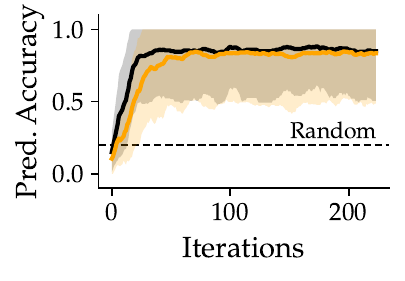}}
	\hspace{0.01cm}
	\subfloat{\includegraphics[width=0.49\linewidth]{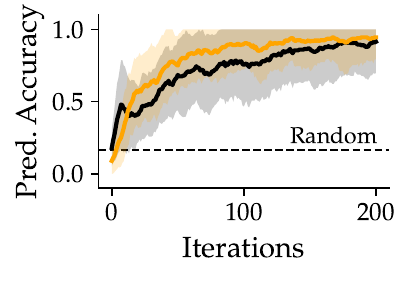}}
    \caption{Prediction accuracy during learning for both contexts in spatially encoded sequence sets $\Seq_3$ (left) and $\Seq_4$ (right). Curves and error bands indicate a moving average ($k=10$) of the mean and one standard deviation.}
    \label{fig:accuracy_pred_last_element}
\end{figure}

Thereby, two context-dependent routing paths through the network are created and allow increased accuracy in predicting the last, context-dependent item (Fig. \ref{fig:accuracy_pred_last_element}). With increasing sequence length, the number of training episodes increases for the network to fully learn the sequences. This is expected, since context needs to be transferred through more overlapping items. For the shorter sequence task $\Seq_3$, prediction accuracy after $200$ iterations of multiple network realizations results in a Bernoulli distribution: while most networks solve the task completely, the rest fail to make correct predictions.

We carry out ablation studies to investigate the network's robustness to noise and sensitivity to hyperparameters using the simplest high-order sequence $\Seq_3$. 

\begin{figure}[htbp]
    \centering
	\subfloat{\includegraphics[width=0.49\linewidth]{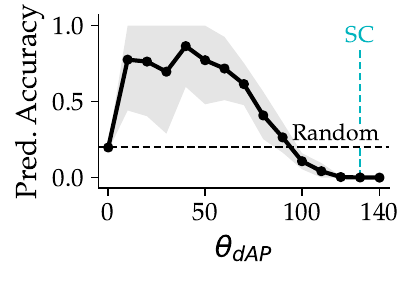}}
	\hspace{0.01cm}
	\subfloat{\includegraphics[width=0.49\linewidth]{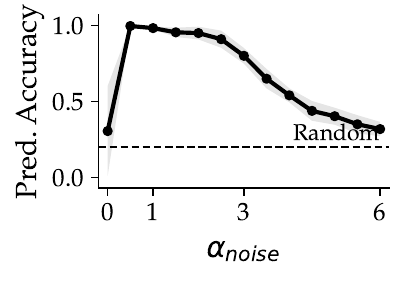}}
    \caption{Prediction accuracy for different ranges of hyperparameters. (Left) The dAP threshold $\theta_\dap$. 
    Thresholds $\theta_\dap > 120\textnormal{pA}$ correspond to a single compartment (SC) neuron model, where neurons do not become predictive. 
    (Right) Firing rate scaling variable $\alpha$. Firing rates of sequence items and background noise are equal if $\alpha = 1$. Default settings are  $\theta_\dap = 20$ and $\alpha = 0.2$. Curves and error bands correspond to the mean and one standard deviation.}
    \label{fig:ablations_excDapThresh_alphaNoise}
\end{figure}

Setting up the dAP threshold $\theta_\dap$ correctly is crucial for learning the sequences (Fig.~\ref{fig:ablations_excDapThresh_alphaNoise}). For a small threshold, too many neurons become predictive and context is lost. The higher the threshold, the fewer neurons are predictive. For a range of $\theta_\dap \in \left[20, 30\right]$, the correct set of neurons switches into the predictive mode for the corresponding transitions. For higher thresholds, few neurons become predictive to carry over context information to the end item, and as such prediction accuracy drops to $0$. This corresponds to a single compartment model since recurrent inputs cannot switch neurons into the predictive state.
The network exhibits robustness over a large range of values of $\alpha$ (Fig.~\ref{fig:ablations_excDapThresh_alphaNoise}). 

\begin{figure}[htbp]
    \centering
	\subfloat{\includegraphics[width=0.49\linewidth]{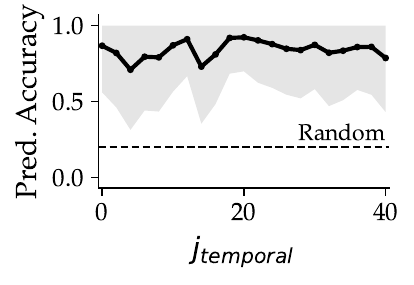}}
	\hspace{0.01cm}
	\subfloat{\includegraphics[width=0.49\linewidth]{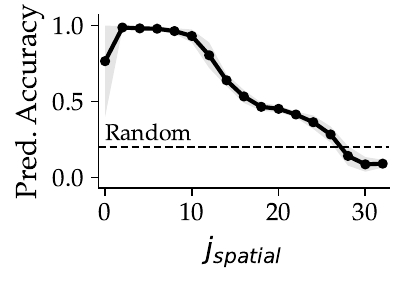}}
    \caption{Prediction accuracy for jittered input patterns. Jitter is applied to the sequence elements during simulation. (Left) Temporal jitter. (Right) Spatial jitter. By default, the network uses a temporal jitter of $j = 1$. Curves and error bands correspond to the mean and one standard deviation.}
    \label{fig:ablations_jitter_temporal_spatial}
\end{figure}

The network is similarly resistant to jitter in the external stimuli (Fig.~\ref{fig:ablations_jitter_temporal_spatial}). Jitter refers to online spike displacements by $j$ steps, either temporally with a step size of $0.1\,\textnormal{ms}$, or spatially, where a step refers to changing the neuron id by $\pm\,j$. While for spatial jitter values of $j > 10$ the network loses the ability to solve high-order transitions, it still correctly predicts the final sequence elements but without taking context into account.
Those results together highlight the usefulness of dendrites as predictive components in a biologically sensible setting.

\subsection{Temporal Encoding}

\begin{figure}[htbp]
    \centering
	\subfloat{\includegraphics[width=0.49\linewidth]{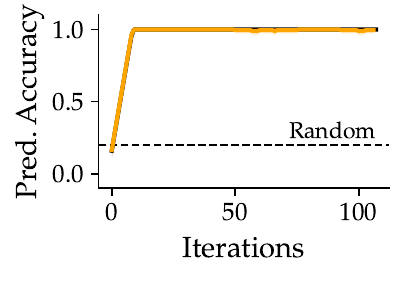}}
	\hspace{0.01cm}
	\subfloat{\includegraphics[width=0.49\linewidth]{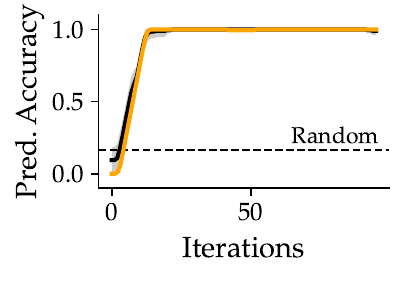}}
    \caption{Prediction accuracy during learning for both contexts in temporally encoded sequences $\Seq_3$ (left) and $\Seq_4$ (right), using precomputed congruent delays $\dEX$. Curves and error bands indicate a moving average ($k=10$) of mean and one standard deviation.}
    \label{fig:temporal_dap_convergence}
\end{figure}

Unlike spatially encoded stimuli, temporal encoding requires learning precise delays $\dEX$, ensuring input spikes arrive nearly synchronously at excitatory neurons. While multiple studies explored learning delays in a supervised fashion \cite{hammouamriLearningDelaysSpiking2023,shrestha_slayer_2018,sun_adaptive_2023}, unsupervised approaches are rarely considered \cite{nadafian_bio-plausible_2020}. 
In this study, we set the delays exactly congruent to the temporal stimuli and assign each excitatory neuron a sparse array of those incoming delays.
Using such artificially calculated delays, temporal patterns allow faster high-order transition learning (Fig.~\ref{fig:temporal_dap_convergence}). During these simulations, networks solve the task also more consistently than with spatial stimuli (compare standard deviations of Fig.~\ref{fig:accuracy_pred_last_element}). Additionally, longer high-order sequences can be considered and solved by the network: while sequences with more than $2$ overlapping elements were solved only occasionally with spatial stimuli, temporal stimuli consistently solved sequences with up to $5$ overlapping elements.

\begin{figure}[!htbp]
        \centering
	\subfloat{\includegraphics[width=0.49\linewidth]{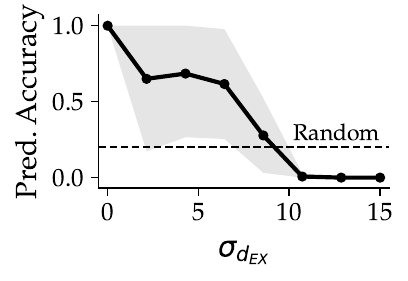}}
	\hspace{0.01cm}
	\subfloat{\includegraphics[width=0.49\linewidth]{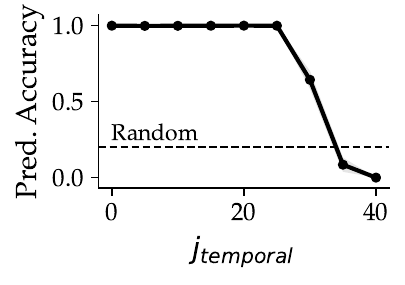}}
    \caption{Prediction accuracy for temporal patterns and congruent delays. (Left) Gaussian noise is applied to the congruent delays, thereby limiting maximum item selectivity. (Right) Temporal jitter is applied to the sequence elements during simulation.}
    \label{fig:temporal_ablation_delays}
\end{figure}

The network performance depends on correct delays. While temporal jittering external spikes does not lead to a decreased performance for up to $2.5\,\textnormal{ms}$ (Fig.~\ref{fig:temporal_ablation_delays}), stimuli are on average congruent to the precomputed delays. In strong contrast, when applying constant noise to the delays, network performance degrades rapidly (Fig.~\ref{fig:temporal_ablation_delays}). This highlights the importance of further research on unsupervised delay learning, since without precomputing delays prediction accuracy would decrease to $0$.

\section{Discussion}
In this study, we extend the work in \cite{bouhadjarSequenceLearningPrediction2022} to allow the processing of sequences of STPs. To this end, we study the robustness of the dAP-based prediction mechanism for predicting sequences in biologically realistic environments. While networks trained on spatial stimuli can detect sequence items, the performance of context-dependent transition learning was observed to be constrained to sequences of low complexity and narrow hyperparameter ranges. Instances of failure are expressed in neurons either not switching into the predictive mode or losing context-dependency during transitions.
We argue that unsupervised delay learning could offer a promising pathway to overcome such limitations since network training is more performant and stable using precomputed delays.

\section{Acknowledgments}
This work was funded by the Federal Ministry of Education, Germany (project NEUROTEC-II grant no. 16ME0398K and 16ME0399).
The authors thank Tom Tetzlaff and Yeshwanth Bethi for valuable discussions on the project.

\bibliographystyle{ieeetr}

\end{document}